\documentclass[aps,prl,preprint,superscriptaddress]{article}

\usepackage[utf8]{inputenc}
\usepackage[left=3cm,right=3cm,top=3cm,bottom=3cm]{geometry}

\usepackage{amsmath}
\usepackage{amsfonts}
\usepackage{amssymb}
\usepackage{mathtools}
\usepackage{braket}
\usepackage{soul}
\usepackage{cite}

\usepackage{graphicx}

\usepackage{color}
\definecolor{nicered}{rgb}{0.7,0.1,0.1}
\definecolor{nicegreen}{rgb}{0.1,0.5,.1}

\usepackage{subcaption}

\usepackage{stackrel}

\usepackage{multirow}

\usepackage{float}

\usepackage{authblk}

\usepackage{comment}

\usepackage{eso-pic}
\newcommand{\reportnum}[2]{
  \AddToShipoutPictureBG*{%
    \AtPageUpperLeft{%
      \hspace{0.75\paperwidth}%
      \raisebox{#1\baselineskip}{%
        \makebox[0pt][l]{\textnormal{#2}}
  }}}%
}

\usepackage[colorlinks=true
,urlcolor=magenta
,anchorcolor=black
,citecolor=blue
,filecolor=black
,linkcolor=red
,menucolor=black
,linktocpage=true
,pdfproducer=medialab
]{hyperref}

\usepackage[compat=1.1.0]{tikz-feynman}

\usepackage{hyperref}
\hypersetup{colorlinks,citecolor=nicegreen,linkcolor=nicered}
\hypersetup{colorlinks=true}

\def\XXint#1#2#3{{\setbox0=\hbox{$#1{#2#3}{\int}$ }
\vcenter{\hbox{$#2#3$ }}\kern-.6\wd0}}

\usepackage{rotating}

\usepackage{multirow}

\usepackage{float}

\usepackage{xcolor}

\usepackage{refcount}

\title{
Implications of two-channel rescattering for charm CP violation
from precise dispersive data}

\author[a]{Antonio~Pich}
\author[b,c]{Eleftheria~Solomonidi}
\author[d]{Luiz~Vale~Silva}

\affil[a]{\it Departament de F\'{i}sica Te\`{o}rica, Instituto de F\'{i}sica Corpuscular,
Universitat de Val\`encia -- Consejo Superior de Investigaciones Cient\'{i}ficas,
Parc Cient\'{i}fic, Catedr\'{a}tico Jos\'{e} Beltr\'{a}n 2, E-46980 Paterna, Valencia, Spain}

\affil[b]{\it PSI Center for Neutron and Muon Sciences, 5232 Villigen PSI, Switzerland}
\affil[c]{\it Physik-Institut, Universit{\"a}t Z{\"u}rich, Winterthurerstrasse 190, 8057 Z{\"u}rich, Switzerland}
\affil[d]{\it Departamento de Matem\'{a}ticas, F\'{i}sica y Ciencias Tecnol\'{o}gicas,

Universidad Cardenal Herrera-CEU, CEU Universities,

46115 Alfara del Patriarca, Val\`{e}ncia, Spain}

\begin{document}
\reportnum{-6}{ZU-TH 30/26}

\maketitle

\begin{abstract}
    
The experimental observation of large CP violation in charm-meson hadronic decays remains theoretically unexplained within the Standard Model. The data-driven approach which accounts for the rescattering between the final-state pion and kaon pairs provides predictions that fall short of the experimental values. However, it relies on a set of inputs that exhibit large uncertainties. In this work, we
make optimal use of the available information on branching ratios and of only one strong-scattering input parameter that is well determined. We find that it remains unlikely to explain the experimental signal within this approach and obtain predictions for sum rules that constrain CP asymmetries of the pion and kaon channels.
\end{abstract}

\section{Introduction}

CP violation is a unique probe of the dynamics of nature at the most fundamental scales.
In the quark sector,
laboratory experiments test the description of CP-violating phenomena, as provided by the Standard Model (SM) of particle physics, via the mixing and decay of bound hadronic states.
A crucial difficulty in decays involving multiple hadrons is dealing with the rescattering among strongly-interacting particles.
One powerful tool to tackle this problem, precursor to the advent of quantum chromodynamics, consists of dispersion relations.
These integral equations require incorporating information about the scattering of stable hadrons.
The Muskhelishvili-Omn\`{e}s matrix is a class of solutions that embodies the effects of final-state rescattering in the inelastic case \cite{Muskhelishvili,Babelon:1976kv,Moussallam:1999aq}.
It is an extension to the Omn\`{e}s factor which applies in the elastic limit \cite{Omnes:1958hv}.
They both have been employed in different electroweak transitions, Higgs \cite{Donoghue:1990xh}, strange- \cite{Pallante:1999qf,Pallante:2000hk,Colangelo:2016ruc},
charm- \cite{Pich:2023kim} and bottom-flavour physics \cite{Heuser:2025mnk,Heuser:2026glv} included, to mention very few examples.

The data-driven method presented in our previous work \cite{Pich:2023kim} uses $\pi \pi$ and $K K$ scattering data to derive the interfering decay amplitudes for $D$-meson decays. Because of the large uncertainties associated with this category of inputs, we utilize the information on $D$-decay branching ratios from all isospin-related channels to narrow down the viable strong input that can correctly predict the measured decays
and, based on the estimated amplitudes, provide a prediction for the CP asymmetry. While the predicted values lie well below the experimental value of the observable $ \Delta A_{CP}^{\rm dir} $ \cite{LHCb:2019hro,LHCb:2022lry}, we would like to address the uncertainties that have not been explicitly displayed in Ref.~\cite{Pich:2023kim}.

When estimating final-state rescattering effects in $D$-meson decays, we need the phase shift manifesting in $\pi\pi \to KK$ scattering.
We employ the parameterization of scattering data \cite{Cohen:1980cq,Etkin:1981sg,Longacre:1986fh} from Refs.~\cite{Pelaez:2018qny,Pelaez:2020gnd}, consistent with basic requirements of analyticity,
unitarity,
and crossing symmetry,
and constrained
by various dispersive relations, which are valid up to about 1.5--1.7~GeV.
Moreover,
their parameterization considers Regge behavior
at energies above about 1.8--2~GeV;
other physical aspects also guide the parameterization that is used to fit the $\pi K \to \pi K$ and $\pi\pi \to KK$ scattering data, such as the existence of cuts, resonances, and Adler zeros at lower energies.
In the case of $\pi\pi \to KK$, such parameterizations are valid up to 2~GeV.

Apart from the uncertainties coming from the dispersive inputs, such as the one of the previous paragraph,
systematic uncertainties stem from restricting our analysis to the two-channel rescattering hypothesis and are difficult to estimate; it has been pointed out that the four-pion channel is important in other decay environments \cite{Ropertz:2018stk}. The four-pion channel would need some modeling and a good determination of the coupling of the source (in this case the weak $D$-meson decay) to this additional channel is important for restricting the number of unknown parameters and improving the predictivity of the model. We have recently verified \cite{Solomonidi:2025cof} that the coupling of $D$ to two intermediate-state hadrons is consistent between the extractions from the $D \to 4 \pi$ and the $D \to \pi \pi \ell^+ \ell^-$ decays, which is an encouraging step and can help us in approximating the third channel that rescatters to $\pi\pi$ and $KK$ as an effective two-body channel. The existence of further isoscalar channels is also possible, such as the six-pion one;
a
partial two-body description of this six-pion channel is an $\omega\omega$ state, which due to the narrow width of the $\omega$ meson would likely appear as a visible cusp in the $\pi\pi$ and $KK$ phases.
The full extension to three channels will be the subject of a future publication. In this work we examine the constraints that arise on the upper limit of the CP asymmetries under the two-channel assumption. These constraints
only rely on one strong phase, the one of $\pi\pi \to KK$ scattering, that is well determined by data as well as the $D$-decay branching ratios.

The role of intermediate resonances in neutral $D$-meson decays to two stable pseudoscalars is discussed, e.g., in Refs.~\cite{Cheng:2010ry,Soni:2019xko,Schacht:2021jaz}.
In the case of CP asymmetries, 
penguin operators mediating the coupling of $D^0$ to $f_0$-states can introduce the required weak phase difference with respect to charged-current operator insertions, while the $f_0$ lineshapes encompass part of the effect of final-state rescattering.
Such strong phase should reflect in the dispersive inputs used in this work through the couplings of those states to pion and kaon pairs.

One might try to employ information from higher-multiplicity $D$ decay modes to extract the coupling induced by charged-current operators of $D^0$ to scalar-isoscalar $f_0$ resonances.
Such approach has been followed in Ref.~\cite{Gronau:1999zt} in order to extract information from Cabibbo-allowed three-body decay modes about the coupling of $D^0$ to $K^\ast_0 (1430)$ and $K^\ast_0 (1950)$, which was
then used in two-body decay predictions.
The method relies on the absence of exotic isospin-$3/2$ resonances, ensuring the smoothness of the soft-pion limit; see also Ref.~\cite{Kou:2023kvp}.
In the present context of singly-Cabibbo-suppressed decay modes, however, isospin-$1$ resonances such as $\pi (1800)$ prevent the same reasoning.
The detailed analysis of three-body decays is beyond the scope of this work; see, e.g., Refs.~\cite{Guimaraes:2014kor,Nakamura:2015qga,Niecknig:2015ija,Niecknig:2017ylb,Achasov:2022hbh,Stamen:2022eda,Escribano:2023zjx,Kou:2023kvp}.

Attempts based on light-cone sum rules cannot like us reproduce the level of CP violation observed by LHCb \cite{Khodjamirian:2017zdu,Lenz:2023rlq}. An extension of their analyses
is currently ongoing.
A flavour symmetry connecting the strange and lighter quarks can be considered in an attempt to obtain a consistent picture across measurements of branching ratios and CP asymmetries \cite{Muller:2015lua,Buccella:2019kpn,Grossman:2019xcj,Gavrilova:2022hbx,Bause:2022jes,Iguro:2024uuw,Gavrilova:2024npn,Fleischer:2025zhl,Gavrilova:2026ryc,Geng:2026qbv}.
Recent fits to data are also presented in Refs.~\cite{Gavrilova:2023fzy,Sinha:2025cuo}.

The study of Cabibbo-allowed and doubly-Cabibbo-suppressed charm-meson decay modes present difficulties of their own. Indeed, to provide a concrete example, $D^+_s \to K^0_S K^+$ receives an annihilation topology contribution that is not suppressed by a small Wilson coefficient. We reserve their analysis to a future work.

This paper is organized as follows.
In Section~\ref{sec:isospin}
we introduce the isospin formalism and discuss relations that follow from isospin invariance.
Unitarity and CPT invariance are used in Section~\ref{sec:unitarity} to establish two-channel rescattering sum rules.
Analyticity is then employed in Section~\ref{sec:analyticity} to implement dispersive inputs in order to discuss the maximum allowed size of CP violation in asymmetries of branching ratios in the case of isospin-0; the interference of isospin-2 or isospin-1 amplitudes with isospin-0 is discussed in Section~\ref{sec:discussion_iso2_iso1}.
Section~\ref{section:NP} briefly discusses how the expressions for the CP asymmetries change in the presence of extra sources of CP violation from a New Physics (NP) sector.
Our conclusions are found in Section~\ref{sec:conclusions}.

\section{Isospin invariance}\label{sec:isospin}

The effective Hamiltonian for the singly-Cabibbo-suppressed charm transitions of interest can be written as

\begin{eqnarray} \label{eq:H_eff}
	&& \mathcal{H}_{\rm eff} = \frac{G_F}{\sqrt{2}} \left[ \, \sum^2_{i = 1} C_i (\mu) \left( \lambda_d Q_i^d + \lambda_s Q_i^s \right)
    - \lambda_b \left( \sum^6_{i = 3} C_i (\mu) Q_i + C_{8g} (\mu) Q_{8g} \right) \right] + \mathrm{h.c.} \,,
\end{eqnarray}
\noindent with
\begin{equation}
	\lambda_q = V^\ast_{c q} V_{u q} \,, \qquad q = d, s, b \,,
\end{equation}
and the basis of operators
\begin{equation}\label{eq:operator_list}
\begin{array}{cc}
\begin{array}{l}
 Q_1^d = ( \overline{d} c )_{V-A} ( \overline{u} d )_{V-A} \,, \\
 Q_2^d = ( \overline{d}_j c_i )_{V-A} ( \overline{u}_i d_j )_{V-A} \,, \\
 Q_1^s = (\overline{s} c)_{V-A} (\overline{u} s)_{V-A} \,, \\
 Q_2^s = (\overline{s}_j c_i)_{V-A} (\overline{u}_i s_j)_{V-A} \,, \\
 \end{array}
&
\begin{array}{l}
 Q_3   = ( \bar{u} c )_{V-A} \sum_q ( \bar{q} q )_{V-A} \,, \\
 Q_4   = ( \bar{u}_j c_i )_{V-A} \sum_q ( \bar{q}_i q_j )_{V-A} \,,\\
 Q_5   = ( \bar{u} c )_{V-A} \sum_q ( \bar{q} q )_{V+A} \,, \\
 Q_6   = ( \bar{u}_j c_i )_{V-A} \sum_q ( \bar{q}_i q_j )_{V+A} \,, \\
  Q_{8g} = - \frac{g_s}{8 \pi^2} m_c \bar{u} \sigma_{\mu \nu} (\mathbf{1} + \gamma_5) G^{\mu \nu} c \,,
\end{array}
\end{array}
\end{equation}
where $ (V - A)_\mu = \gamma_\mu (\mathbf{1} - \gamma_5) $, $i, j$ are colour indices, and $\mu$
is the renormalization scale.

The decay amplitudes to final states of specific charge (``flavour basis") can also be expressed via the use of the Wigner-Eckart theorem as linear combinations of amplitudes to isospin-specific final states. We parameterize them as follows: 
\begin{eqnarray}
A[D^0\to\pi^0\pi^0] &\! \! = &\! \! - \frac{1}{\sqrt{6}}\; T^0_{\pi\pi} + \frac{1}{\sqrt{3}}\; T^2_{\pi\pi} \, ,
\nonumber\\
A[D^0\to\pi^+\pi^-]&\! \!\equiv &\! \! \frac{1}{\sqrt{2}}\, A[D^0\to\frac{1}{\sqrt{2}}\, (\pi^+\pi^-+\pi^-\pi^+)] \, = \,  -\frac{1}{\sqrt{6}}\; T^0_{\pi\pi} -  \frac{1}{2\sqrt{3}}\; T^2_{\pi\pi} \, ,
\nonumber\\
A[D^+\to\pi^+\pi^0] &\! \!\equiv &\! \! \frac{1}{\sqrt{2}}\, A[D^+\to\frac{1}{\sqrt{2}}\, (\pi^+\pi^0+\pi^0\pi^+)] \, = \,  \frac{\sqrt{3}}{2\sqrt{2}}\; T^2_{\pi\pi} \, ,
\end{eqnarray}

\begin{eqnarray}
A[D^0\to K^+ K^-] &=& \frac{1}{2} \left( T^{11}_{KK}+T^{13}_{KK} - T^0_{KK}\right) \, ,
\nonumber\\
A[D^0\to K^0 \overline{K}^0] &=& \frac{1}{2}\, \left( -T^{11}_{KK} -T^{13}_{KK} - T^0_{KK}\right) \, ,
\nonumber\\
A[D^+\to K^+ \overline{K}^0] &=& T^{11}_{KK}-\frac{1}{2}T^{13}_{KK} \, .
\end{eqnarray}

The superscripts 0, 1, 2 indicate the final-state isospin, while the second superscript for the $KK$ amplitudes indicates the isospin of the acting operator, namely 1 and 3 stand for $\Delta I=1/2$ and $3/2$ respectively. 
The isospin-invariant amplitudes can be split into amplitudes of different weak phases, generically written as 
\begin{equation}\label{eq:intro_tilde_amplitudes}
    T_{\pi\pi}^{I}=\lambda_d \Tilde{A}_{d,\pi}^{(I)}+\lambda_s \Tilde{A}_{s,\pi}^{(I)}\,, \qquad T_{KK}^{I}=\lambda_d \Tilde{A}_{d,K}^{(I)}+\lambda_s \Tilde{A}_{s,K}^{(I)}\,,
\end{equation}
wherein we have used the CKM unitarity to express the $\lambda_b$ coefficient as (minus) the sum of the other two. 
Based on the isospin decomposition and taking into account that the $\Tilde{A}$ amplitudes contain CP-even strong phases, one can then express the CP asymmetries as follows:

\begin{eqnarray}
\label{acppi2and0}
  A_{CP}(D^+\rightarrow \pi^+\pi^0)& =& \frac{|A[D^0\to\pi^+\pi^-]|^2}{|A[D^0\to\pi^+\pi^0]|^2} \;\frac{9}{2}\, a_{CP}(\pi\pi[2/2]) \,, \label{acppi+pi0} \\
  A_{CP}(D^0\rightarrow \pi^+\pi^-)& =& a_{CP}(\pi\pi[0/0])+ a_{CP}(\pi\pi[2/0])+a_{CP}(\pi\pi[2/2]) \,, \\
  A_{CP}(D^0\rightarrow \pi^0\pi^0) &=& \frac{|A[D^0\to\pi^+\pi^-]|^2}{|A[D^0\to\pi^0\pi^0]|^2} \, \left( a_{CP}(\pi\pi[0/0])-2\, a_{CP}(\pi\pi[2/0])+4\, a_{CP}(\pi\pi[2/2])\right) ,\quad
\end{eqnarray}
where we have defined 
\begin{eqnarray}
  a_{CP}(\pi\pi[0/0])& =&\frac{1}{3}\,\frac{\mathrm{Im}(\lambda_s \lambda_d^*)\,\mathrm{Im}\left(\Tilde{A}_{d,\pi}^{(0)}\,(\Tilde{A}_{s,\pi}^{(0)})^*\right)}{|A[D^0\to\pi^+\pi^-]|^2}\, , \\
  a_{CP}(\pi\pi[2/0])& =&\frac{\sqrt{2}}{6}\,\frac{\mathrm{Im}(\lambda_s \lambda_d^*)\left(\mathrm{Im}\left(\Tilde{A}_{d,\pi}^{(2)}\,(\Tilde{A}_{s,\pi}^{(0)})^*\right)+\mathrm{Im}\left(\Tilde{A}_{d,\pi}^{(0)}\,(\Tilde{A}_{s,\pi}^{(2)})^*\right)\right)}{|A[D^0\to\pi^+\pi^-]|^2}\,, \\
  a_{CP}(\pi\pi[2/2])& =&\frac{1}{6}\,\frac{\mathrm{Im}(\lambda_s \lambda_d^*)\,\mathrm{Im}\left(\Tilde{A}_{d,\pi}^{(2)}\,(\Tilde{A}_{s,\pi}^{(2)})^*\right)}{|A[D^0\to\pi^+\pi^-]|^2}\, ,
\end{eqnarray}
the asymmetries $[I/J]$ which come from the interference of isospin-$I$ and isospin-$J$ amplitudes with different weak phases.
Analogous expressions hold for the kaon-pair modes.
In all the above equations and in the ones to follow, we denote as $|A[D\to PP']|^2$ as well as $B_{\{+0,00,+-\}}$ the CP-averaged squared magnitudes of the amplitudes and the CP-averaged branching ratios, given that CP asymmetries are very small.
Throughout this work, CP asymmetries take into account direct CP violation only.

The above equations lead to expressing a CP-violating observable for the flavour-specific CP asymmetries, defined analogously to Belle~II \cite{Belle-II:2025rmf}, as follows \cite{Grossman:2012eb,HFLAV:2022pwe,Wang:2022nbm}:

\begin{equation}
      \begin{aligned}
    R (0/0) \, &\equiv\, \frac{A_{CP}(D^0\to\pi^+\pi^-)}{1 + \frac{\tau_{D^0}}{B_{+-}}\big(\frac{B_{00}}{\tau_{D^0}} - \frac{2}{3}\frac{B_{+0}}{\tau_{D^+}} \big)
    }
    +
    \frac{A_{CP}(D^0\to\pi^0\pi^0)}{1 + \frac{\tau_{D^0}}{B_{00}}\big(\frac{B_{+-}}{\tau_{D^0}} - \frac{2}{3}\frac{B_{+0}}{\tau_{D^+}} \big)}+\frac{A_{CP}(D^+\to\pi^+\pi^0)}{1 -\frac{3}{2}\frac{\tau_{D^+}}{\tau_{D^0}}\frac{ B_{00} +B_{+-}}{ B_{+0}}} \\
&=\, 2\,\frac{\mathrm{Im}(\lambda_s \lambda_d^*)\,\mathrm{Im}\left(\Tilde{A}_{d,\pi}^{(0)}\,(\Tilde{A}_{s,\pi}^{(0)})^*\right)}{|T_{\pi\pi}^0|^2} \,. &
    \end{aligned}
\end{equation}

\noindent
This relation
offers an observable that isolates the CP asymmetries originating in the interference of isospin-$0$ amplitudes with themselves.
In a later section we will present our prediction for this observable under the assumption of two-channel final-state rescattering.
Despite the claim made in Ref.~\cite{Belle-II:2025rmf}, a zero value for this observable, while having a non-zero CP asymmetry, such as in $D^0\to\pi^+\pi^-$ \cite{LHCb:2019hro,LHCb:2022lry}, does not necessarily point at beyond-the-SM dynamics.
This is so because the following observable $R (2/0)$, isolating isospin 2/0-induced CP asymmetry, can be non-zero:
\begin{equation}
    \begin{aligned}
    R (2/0)\,  &\equiv\, \frac{A_{CP}(D^0\to\pi^+\pi^-)}{1 + \frac{\tau_{D^0}}{B_{+-}}\big( -2 \frac{B_{00}}{\tau_{D^0}} + \frac{2}{3}\frac{B_{+0}}{\tau_{D^+}} \big)
    }
    +
    \frac{A_{CP}(D^0\to\pi^0\pi^0)}{1 - \frac{1}{2} \frac{\tau_{D^0}}{B_{00}}\big(\frac{B_{+-}}{\tau_{D^0}} + \frac{2}{3}\frac{B_{+0}}{\tau_{D^+}} \big)}+\frac{A_{CP}(D^+\to\pi^+\pi^0)}{1 +\frac{3}{2}\frac{\tau_{D^+}}{\tau_{D^0}}\frac{ -2 B_{00} +B_{+-}}{ B_{+0}}} \\
&=\, 2\, \frac{\mathrm{Im}(\lambda_s \lambda_d^*)\left(\mathrm{Im}\left(\Tilde{A}_{d,\pi}^{(2)}\,(\Tilde{A}_{s,\pi}^{(0)})^*\right)+\mathrm{Im}\left(\Tilde{A}_{d,\pi}^{(0)}\,(\Tilde{A}_{s,\pi}^{(2)})^*\right)\right)}{T_{\pi\pi}^0 (T_{\pi\pi}^2)^\ast + T_{\pi\pi}^2 (T_{\pi\pi}^0)^\ast} \,, &
    \end{aligned}
    \label{eq: R20}
\end{equation}
and of SM origin.
It is clear that the mode $D^+\rightarrow \pi^+\pi^0$ probes isospin 2/2 interference.
It is not possible to write analogous relations for the kaon pair decay modes, due to the presence of both $ T^{11}_{KK} $ and $ T^{13}_{KK} $ amplitudes, which moreover can in general carry different strong phases, together with $ T^0_{KK} $.

By inspecting the quark composition of the operators in Eq.~\eqref{eq:operator_list}, one can see that potential contributions are
\begin{eqnarray}
    && T_{\pi\pi}^2=\lambda_d \Tilde{A}_{d,\pi}^{(2)}\,, \label{Tpipi2} \\
    && T_{\pi\pi}^0=\lambda_d \Tilde{A}_{d,\pi}^{(0)}+\lambda_s \Tilde{A}_{s,\pi}^{(0)}\,, \label{twoamplitudesI0}\\
    && T_{KK}^{13}=\lambda_d \Tilde{A}_{d,K}^{(13)}\,,\\
    && T_{KK}^{11}=\lambda_d \Tilde{A}_{d,K}^{(11)}+\lambda_s \Tilde{A}_{s,K}^{(11)}\,,\\
    && T_{KK}^{0}=\lambda_d \Tilde{A}_{d,K}^{(0)}+\lambda_s \Tilde{A}_{s,K}^{(0)}\,. \label{twoamplitudesI0_kaons}
\end{eqnarray}
Indeed, the charged-current operators $Q_1^s$ and $Q_2^s$, as well as the gluonic penguin and dipole operators, cannot change isospin number by $3/2$, while isospin number is preserved by QCD interactions. More generally then, and following the above notation, final states of isospin-2 and the amplitudes mediated by the operator component that changes isospin by $3/2$ do not receive contributions proportional to $\lambda_s$.

The definitions for $a_{CP}(\pi\pi[2/0])$ and $a_{CP}(\pi\pi[2/2])$ simplify under the later expression for $T_{\pi\pi}^2$.
As a first remark, the CP asymmetry of the charged $D$-meson decay mode to two pions vanishes in the SM at the current level of precision and thus constitutes by itself a null test of the SM.\footnote{Potential SM CP-violating effects in this channel arise from the highly GIM-suppressed electroweak penguin operators, or at higher order in $G_F \cdot \alpha_{\mathrm{em}}$, negligible at the current level of precision and ignored throughout this work. For discussions, see Refs.~\cite{Cirigliano:2003gt,Grossman:2026qew}.} Accordingly, in the observable $R(2/0)$, the second term in the numerator of the second line of Eq.~\eqref{eq: R20} vanishes.

\section{Unitarity, CPT invariance and isospin-0 sum rules}\label{sec:unitarity}

In this section we discuss the results generically obtained from unitarity and CPT invariance. We focus on the final states with isospin zero, as they present inelastic rescattering. As in our previous work \cite{Pich:2023kim}, we assume that unitarity holds in the restricted $\{\pi\pi, KK\}$ subspace. 

Unitarity of the total $S$-matrix to lowest order in electroweak interactions leads
to \cite{Babelon:1976kv,Franco:2012ck}
\begin{equation}\label{eq:unitarity_main}
	\Sigma^{1/2} \, \begin{pmatrix}
	\mathcal{C}^\pi_0 \\
	\mathcal{C}^K_0 \\
	\end{pmatrix} \, = \,
	(S_S)^{I=0}_{J=0} \;\, \Sigma^{1/2} \,
	\begin{pmatrix}
	(\mathcal{C}^\pi_0)^\ast \\
	(\mathcal{C}^K_0)^\ast \\
	\end{pmatrix} \,, \quad  \mathcal{C} = \mathcal{A}, \mathcal{B} \,,
\end{equation}
where $S_S$ is the partial-wave-projected strong-rescattering submatrix, and the kinematic factor (hereafter calculated at $ s = M_D^2 $) is

\begin{equation}
	\Sigma (s) = {\rm diag} \!\left[ \Theta (s - 4 M^2_\pi)\,\sigma_\pi (s)\, , \,\Theta (s - 4 M^2_K)\,\sigma_K (s) \right] \,, \quad \sigma_i (s) = \sqrt{1 - 4 M^2_i / s} \,.
\end{equation}
This equation is displayed in the particular isosinglet-scalar case (i.e., isospin equal to zero, $I=0$, and in the $S$-wave $J=0$), although other cases are also possible in general.
CPT invariance has been employed in deriving Eq.~\eqref{eq:unitarity_main} (by equating the time-reversed $PP \to D$ amplitudes to the CP-conjugate of $D\to PP$),
and $ \mathcal{C} = \mathcal{A} $ or $ \mathcal{B} $ denote the CP-even or CP-odd amplitudes, respectively, for $D \to \pi \pi$ and $K K$ decays. We have used the even and odd notation for the sake of clarity, although this splitting is not rephasing-invariant; to see the correspondence, in the case of the $T_{\pi\pi}^0$ of the previous section one would have $\mathcal{A}^\pi_0 \equiv \mathrm{Re} T_{\pi\pi}^0=\lambda_d \,\Tilde{A}_{d,\pi}^{(0)}+\mathrm{Re}\lambda_s\,\Tilde{A}_{s,\pi}^{(0)}$ and $\mathcal{B}^\pi_0 \equiv \mathrm{Im} T_{\pi\pi}^0=\mathrm{Im}\lambda_s\, \Tilde{A}_{s,\pi}^{(0)}$ if $\lambda_d$ is chosen to be real. Eq.~\eqref{eq:unitarity_main} in the elastic limit establishes Watson's theorem.

Using the unitarity relation~\eqref{eq:unitarity_main} for both CP-even and CP-odd amplitudes, along with the unitarity of the strong $S_S$ submatrix, one gets a so-called CPT-unitarity relation \cite{Weinberg:1995mt,Bigi:2000yz,Soni:2019xko},
\begin{equation}\label{eq:CPT}
    \sigma_\pi \, | \mathcal{A}^\pi_0 | \, | \mathcal{B}^\pi_0 | \, \sin \left( \arg ( \mathcal{A}^\pi_0 ) - \arg ( \mathcal{B}^\pi_0 ) \right) + \sigma_K \, | \mathcal{A}^K_0 | \, | \mathcal{B}^K_0 | \, \sin \left( \arg ( \mathcal{A}^K_0 ) - \arg ( \mathcal{B}^K_0 ) \right) = 0 \,,
\end{equation}
which will establish a more direct relation between different CP asymmetries from amplitudes of isospin-$0$.
In terms of the amplitudes introduced in Eq.~\eqref{eq:intro_tilde_amplitudes} we then have
\begin{equation}\label{eq:CPT_tilde}
    \sigma_\pi \, | \Tilde{A}_{d,\pi}^{(0)} | \, | \Tilde{A}_{s,\pi}^{(0)} | \, \sin \left( \arg (\Tilde{A}_{d,\pi}^{(0)}) - \arg (\Tilde{A}_{s,\pi}^{(0)}) \right) + \sigma_K \, | \Tilde{A}_{d,K}^{(0)} | \, | \Tilde{A}_{s,K}^{(0)} | \, \sin \left( \arg (\Tilde{A}_{d,K}^{(0)}) - \arg (\Tilde{A}_{s,K}^{(0)}) \right) = 0 \,.
\end{equation}
As will be illustrated in the following section, as a consequence of inelasticity the phases of the CP-even and CP-odd amplitudes $\mathcal{A}^\pi_0$ and $\mathcal{B}^\pi_0$ (and of $\mathcal{A}^K_0$ and $\mathcal{B}^K_0$), or equivalently $\Tilde{A}_{d,\pi}^{(0)}$ and $\Tilde{A}_{s,\pi}^{(0)}$ (or $\Tilde{A}_{d,K}^{(0)}$ and $\Tilde{A}_{s,K}^{(0)}$), can be different \cite{Franco:2012ck}; note that such a possibility is not available in the context of direct CP violation in kaon meson decays.
The CPT-unitarity relation can then be expressed in terms of the 0/0 CP asymmetries defined before as 
\begin{eqnarray}
    \frac{a_{CP}({\pi\pi}[0/0])}{a_{CP}({KK}[0/0])}\, =\, -\frac{2}
    {3}\,\frac{|A[D^0\to K^+K^-]|^2}{|A[D^0\to\pi^+\pi^-]|^2}\,\frac{\sigma_K}{\sigma_\pi} \,.
\end{eqnarray}

A third $I=0$ channel would generically modify the relation of Eq.~\eqref{eq:unitarity_main} and subsequently the CPT-unitarity relations of Eqs.~\eqref{eq:CPT} and \eqref{eq:CPT_tilde}, appending to the sum rule an additional term proportional to the CP asymmetry of that channel; see, e.g., Refs.~\cite{Moussallam:1999aq,Franco:2012ck}.

\section{Analyticity and evaluation of the CP asymmetries}\label{sec:analyticity}

\begin{figure*}[t]
    \centering
    \includegraphics[scale=0.45]{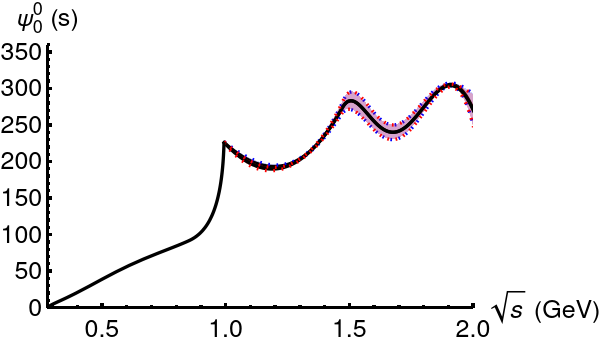}
    \caption{
    The $ \psi^0_0 (s) $ phase, given in degrees, for solutions~B (blue) \cite{Longacre:1986fh} and C (red) \cite{Cohen:1980cq,Etkin:1981sg}, which are very compatible; below the kaon pair threshold, $ \psi^0_0 (s) $ is the phase extracted from $\pi \pi$ scattering.
    Similar figures have already been presented in Ref.~\cite{Pich:2023kim}, where additional details can be found.
    }
    \label{fig:phase_shift}
\end{figure*}

In Ref.~\cite{Pich:2023kim} we factorized the strong rescattering from the weak transition amplitude through the introduction of the Muskhelishvili-Omn\`{e}s matrix $ \Omega^{(0)} $, thus writing
\begin{equation}\label{eq:Omnesfactorisation}
	\begin{pmatrix} T_{\pi \pi}^0
\\ T_{K K}^0 \end{pmatrix}
 \, = \,  \Omega^{(0)} \; 
 \begin{pmatrix} \lambda_d \, T^{CC}_{\pi \pi} - \lambda_b \, T^{P}_{\pi \pi}
  \\ \lambda_s \, T^{CC}_{K K} - \lambda_b \, T^{P}_{K K} \end{pmatrix} = \,  \Omega^{(0)} \; 
 \begin{pmatrix} \lambda_d \, ( T^{CC}_{\pi \pi} + T^{P}_{\pi \pi} ) + \lambda_s \, T^{P}_{\pi \pi}
  \\ \lambda_d \, T^{P}_{K K} + \lambda_s \, ( T^{CC}_{K K} + T^{P}_{K K} ) \end{pmatrix} \, .
\end{equation}
The relation of the bare amplitudes $ T^{CC}_{\pi \pi, K K} $ and $ T^{P}_{\pi \pi, K K} $ with the amplitudes introduced in Eq.~\eqref{eq:intro_tilde_amplitudes} can be directly read from the last identity in Eq.~\eqref{eq:Omnesfactorisation}. They depend linearly on the Wilson coefficients of the dimension-6 contact interactions induced by the electroweak dynamics (``$CC$'' and ``$P$'' stand for charged-current and penguin operators, respectively), can be determined to leading order in the large-$N_C$ counting, and expressed in terms of meson decay constants and meson-to-meson form factors.
The matrix $ \Omega^{(0)} $ collects instead the sub-leading effects in the large-$N_C$ counting responsible for the rescattering between pairs of pions and kaons, and is calculated as the numerical solution to an inelastic dispersive integral. In the two-channel coupled analysis,
\begin{equation}
    \Omega^{(0)} =
    \begin{pmatrix}
        | \Omega^{(0)}_{11} |\, e^{i \, \phi_{11}} & | \Omega^{(0)}_{12} |\, e^{i \, \phi_{12}} \\
        | \Omega^{(0)}_{21} |\, e^{i \, \phi_{21}} & | \Omega^{(0)}_{22} | \, e^{i \, \phi_{22}} \\
    \end{pmatrix} \,.
\end{equation}

In this framework, we can express the isospin 0/0 CP asymmetries introduced before as 
\begin{equation}
 a_{CP}(\pi\pi[0/0])\, =\, \frac{\omega^{({\rm Im})}_\pi \mathcal{J}}{3\, |A[D^0\rightarrow \pi^+\pi^-]|^2}\,\bigl(T^{CC}_{\pi \pi}T^{CC}_{K K} +T^{P}_{\pi \pi}T^{CC}_{K K} \, + T^{CC}_{\pi \pi}T^{P}_{K K}\bigr) \,,
\end{equation}
\begin{equation}
 a_{CP}(KK[0/0])\, =\, \frac{\omega^{({\rm Im})}_K \mathcal{J}}{2\, |A[D^0\rightarrow K^+K^-]|^2}\,\bigl(T^{CC}_{\pi \pi}T^{CC}_{K K} +T^{P}_{\pi \pi}T^{CC}_{K K}+T^{CC}_{\pi \pi}T^{P}_{K K}\bigr) \,,
\end{equation}
where $\mathcal{J}=\text{Im}(\lambda_s \lambda_b^*)=\text{Im}(\lambda_b \lambda_d^*)=\text{Im}(\lambda_d \lambda_s^*)$ is the Jarlskog invariant. The above CP asymmetries depend on numerators proportional to $ \omega^{({\rm Im})}_\pi $ and $ \omega^{({\rm Im})}_K $ as the only parameters determined by the dispersive analysis, where 
\begin{align}
    \omega^{({\rm Im})}_\pi = \text{Im} \{ \Omega^{(0) \ast}_{1 1} \Omega^{(0)}_{1 2} \} = - | \Omega^{(0)}_{11} | \, | \Omega^{(0)}_{12} | \, \sin (\delta_1) \,, \\
    \omega^{({\rm Im})}_K = \text{Im}\{\Omega_{21}^{(0) \ast} \Omega_{22}^{(0)}  \} = - | \Omega^{(0)}_{21} | \, | \Omega^{(0)}_{22} | \, \sin (\delta_2) \,.
\end{align}
These quantities encode the non-vanishing differences of strong phases $ \delta_i = \phi_{i1} - \phi_{i2} $, $ i = 1, 2$, necessary for CP violation.
Such quantities, defined exclusively from isospin-$0$ final-state modes, can be different from zero only in the inelastic case, since otherwise $| \Omega^{(0)}_{12} |=| \Omega^{(0)}_{21} |=0$.

Using Eq.~\eqref{eq:Omnesfactorisation} and the CPT-unitarity relation of Eq.~\eqref{eq:CPT} one gets an equation already presented in our previous work \cite{Pich:2023kim}:
\begin{equation}\label{eq:sum_rule_rescattering}
    {\rm Im} \{ \Omega^{(0) \dagger} \, \Sigma \, \Omega^{(0)} \} = 0 \quad \Leftrightarrow \quad \sigma_\pi \, \omega^{({\rm Im})}_\pi + \sigma_K \, \omega^{({\rm Im})}_K = 0 \,.
\end{equation}
Since $\sigma_\pi$ and $\sigma_K$ are both positive, Eq.~\eqref{eq:sum_rule_rescattering} means that $\omega^{({\rm Im})}_\pi$ and $\omega^{({\rm Im})}_K$, as well as $a_{CP}(\pi\pi[0/0])$ and $a_{CP}(KK[0/0])$, carry opposite signs, which is beneficial for the difference of CP asymmetries in pion and kaon pairs.

Our goal now is determining how large the CP asymmetries proportional to $\omega^{({\rm Im})}_\pi$ and $ \omega^{({\rm Im})}_K $ can be.
For this sake,
note that the values of experimental
branching ratios fix $|T_{\pi\pi}^0|$ and $| T_{KK}^0 |$ to a very good accuracy. Moreover, as per our previous work \cite{Pich:2023kim},
the quantity $ \det \Omega^{(0)} $ is fixed by a single rescattering phase $ \psi^0_0 (s) $ under the two-channel hypothesis.
Consider then
the following expressions\footnote{At this stage we do not discuss $ \arg \det \Omega^{(0)} $ as it would introduce a dependence on yet another angle.}

\begin{equation}\label{eq:A0piSquared_constraint}
    | T_{\pi\pi}^0 |^2 = |\lambda_d|^2\,(T_{\pi \pi}^{CC})^2 \, | \Omega^{(0)}_{11} |^2 + |\lambda_s|^2\,(T_{K K}^{CC})^2 \, | \Omega^{(0)}_{12} |^2 + 2 \,\mathrm{Re}\left(\lambda_d \lambda_s^*\right) T_{\pi \pi}^{CC} \, T_{K K}^{CC} \, | \Omega^{(0)}_{11} | \, | \Omega^{(0)}_{12} | \, \cos ( \delta_1 ) \,,
\end{equation}

\begin{equation}\label{eq:A0KSquared_constraint}
    | T_{KK}^0 |^2 = |\lambda_d|^2\,(T_{\pi \pi}^{CC})^2 \, | \Omega^{(0)}_{21} |^2 +  |\lambda_s|^2\,(T_{K K}^{CC})^2\, | \Omega^{(0)}_{22} |^2 + 2 \,\mathrm{Re}\left(\lambda_d \lambda_s^*\right) T_{\pi \pi}^{CC} \, T_{K K}^{CC} \, | \Omega^{(0)}_{21} | \, | \Omega^{(0)}_{22} | \, \cos ( \delta_2 ) \,,
\end{equation}

\begin{equation}\label{eq:DetOmega2_constraint}
    | \det \Omega^{(0)} |^2 = | \Omega^{(0)}_{11} |^2 \, | \Omega^{(0)}_{22} |^2 + | \Omega^{(0)}_{12} |^2 \, | \Omega^{(0)}_{21} |^2 - 2 \, | \Omega^{(0)}_{11} | \, | \Omega^{(0)}_{22} | \, | \Omega^{(0)}_{12} | \, | \Omega^{(0)}_{21} | \, \cos ( \delta_1 - \delta_2 ) \,.
\end{equation}
Obviously, $ \sin ( \delta_i )^2 + \cos ( \delta_i )^2 = 1 $, $ i = 1, 2 $; also, $ | \Omega^{(0)}_{i j} | \geq 0 $, $ i, j = 1, 2 $.
The other terms arising from Eq.~\eqref{eq:Omnesfactorisation}, namely the penguin-operator matrix elements multiplied by the tiny $\lambda_b$ and, consistently, terms carrying the weak phase $\mathrm{Im}\left(\lambda_d \lambda_s^*\right)$, are only important for the CP asymmetries and are safely omitted in the above expressions. The phase $ \psi^0_0 (s) $, needed to determine $ \det \Omega^{(0)} $, describes
$ \pi \pi \to K K $
for isospin-$0$ and in the $S$-wave above the kaon pair threshold.
There is no clear indication of inelasticity
in the region below the kaon pair threshold (which is then influenced only by the effect of virtual kaons). Therefore, one has that $ \psi^0_0 (s) = \delta^0_0 (s) $ in the latter region, following from Watson's theorem, where $\delta^0_0 (s)$ is the phase extracted
from $ \pi \pi \to \pi \pi $.
Their profiles are found in Fig.~1 of Ref.~\cite{Pich:2023kim}, which is reproduced here for convenience in Fig.~\ref{fig:phase_shift}.

In contrast to the full $ \Omega^{(0)} (s) $ matrix, the hadronic quantity
$ \det \Omega^{(0)} (s) $ admits an expression in closed form.
Since the phase shift $\psi^0_0(s)$ goes to a constant at large energies,
$ \det \Omega^{(0)} (s) $ 
has a once-subtracted Omn\`{e}s solution \cite{Omnes:1958hv,Moussallam:1999aq,Pich:2023kim}:
\begin{equation}
 \text{det}\Omega^{(0)} (s)
 =\exp\{i \psi^0_0(s)\}\exp\left\{ \frac{s-s_0}{\pi} PV \int_{4M_{\pi}^2}^{\infty} dz \frac{\psi^0_0(z)}{(z-s_0)(z-s)} \right\} \,,
 \label{analyticalsol}
\end{equation}
where $PV$ stands for the principal value.
This solution satisfies $ \text{det}\Omega^{(0)} (s_0) = 1 $ by construction (where $s_0 = M^2_\pi$ here), while the subtraction constant
of the
dispersion relation for $\Omega^{(0)} (s)$
consists of the bare amplitudes previously presented, which in particular carry the source $\mathcal{J}$ of CP violation.
A larger number of subtractions is possible, but this requires the introduction of a larger number of subtraction constants.
The expression of the Omn\`{e}s solution for a different number of subtractions is found, e.g., in Ref.~\cite{Pallante:2000hk}.
In physical terms, the weaker dependence on $\psi^0_0 (s)$ at high energies for a larger number of subtractions is exchanged by the knowledge of more subtraction constants, which are thus low-energy constants encoding the ultraviolet behavior of rescattering.

\begin{figure}
    \centering
    \begin{minipage}{0.48\textwidth}
    \centering
    \includegraphics[scale=0.55]{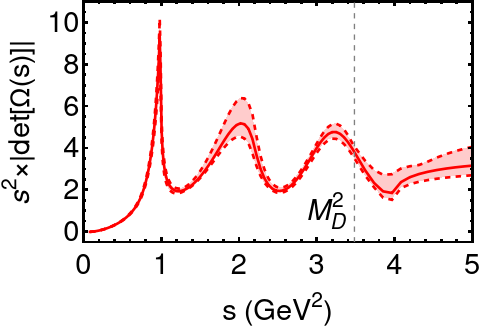}
    \end{minipage}
    \hfill
    \begin{minipage}{0.48\textwidth}
    \vspace{-1cm}
    \centering
    \includegraphics[scale=0.44]{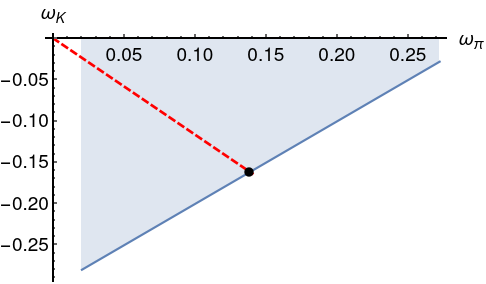}
    \end{minipage}    
    \caption{
    (Left) Explicit analytical solution of $s^2$ times $ | \det \Omega^{(0)} (s) | $. The middle solid red line corresponds to the nominal strong phase $\psi^0_0 (s)$. The band covers the solutions when the strong phase is varied within its uncertainties.
    (Right) Allowed region in the $ \omega^{({\rm Im})}_\pi $ vs. $ \omega^{({\rm Im})}_K $ plane when $ \omega^{({\rm Im})}_\pi $ and $ \omega^{({\rm Im})}_K $ have opposite signs and $ \omega^{({\rm Im})}_K $ is negative. The blue boundary represents the limit of Eq.~\eqref{eq:analytical_bound}, the red dashed line the sum rule of Eq.~\eqref{eq:sum_rule_rescattering}, and the black dot the situation of maximum isospin-$0$ CP asymmetries for a fixed dispersive input $\psi^0_0 (s)$.}
    \label{fig:omegapi_omegaK_opposite_signs}
\end{figure}

The $ \det \Omega^{(0)} (s) $ profile is displayed in
the left panel of
Fig.~\ref{fig:omegapi_omegaK_opposite_signs}.\footnote{While the numerical calculation of the integral in Eq.~\eqref{analyticalsol} is straightforward in this case, a
qualitative understanding can be drawn from the representation of the phase shift $\psi^0_0 (s)$ by means of a toy composition of arctangent functions inspired by Breit-Wigner lineshapes (or step-functions in the limit of no width) and Gaussian bumps, which admit explicit, closed-form expressions for their Hilbert transforms. Dynamical approximations to the Omn\`{e}s factor, not employed here, are discussed, e.g., in Ref.~\cite{Barton:1965}.}
The peak seen at about $1~\text{GeV}^2$ is due to the $f_0 (980)$ resonance, while the smaller peaks at about $2~\text{GeV}^2$ and $3.2~\text{GeV}^2$ are due to the bumps seen in $ \psi^0_0 (s) $ at about 1.5~GeV and 1.9~GeV, respectively.
Such bumps are possibly due to the presence of other scalar-isoscalar resonances.
Since $ \det \Omega^{(0)} (s) \to s^x $ for large energies, as can be verified from Eq.~\eqref{analyticalsol}, with $ x = - (\psi^0_0 (+\infty) - \psi^0_0 (4 M_\pi^2))/\pi $,
the overall damping of $\det \Omega^{(0)} (s)$ is due to the phase-shift difference $ \psi^0_0 (+\infty) - \psi^0_0 (4 M_\pi^2) $.
The index $x$ that applies for the determinant is related to the individual indices $x_i$, $i=1,2$, governing the asymptotic behaviors of the two fundamental solutions of the coupled dispersive integrals as $x=x_1 + x_2$, i.e., the amplitudes for the pion and kaon decay modes result from linear combinations (with polynomial coefficients) of functions with asymptotic behaviors bounded by $s^{x_1}$ and $s^{x_2}$ \cite{Muskhelishvili,Babelon:1976kv,Pich:2023kim}.

The phase $\psi^0_0 (+\infty)$ is not known experimentally: $\psi^0_0$ at 2~GeV is known from data to be in the range $250^\circ$--$290^\circ$, and we extrapolate from $\psi^0_0 (2~\text{GeV}^2)$ to a value of $\psi^0_0 (+\infty)$ equal to
a multiple of $180^\circ$, namely $360^\circ$.
A higher (lower) value of $\psi^0_0 (+\infty)$ would produce a stronger (weaker, respectively) damping.
In particular,
considering $180^\circ$ instead of $360^\circ$ could jeopardize the asymptotic behavior of the individual isospin-$0$ $\pi \pi$ and $K K$ form factors \cite{Lepage:1980fj,Leutwyler:2002hm}, since in a two-channel analysis one of them would then forcefully go asymptotically to a non-vanishing constant instead of decreasing with a power of the energy.
On the other hand,
choosing a larger integer multiple of $180^\circ$ (i.e., $540^\circ$, etc.) would consist of demanding a strong dynamical change of the phase shift at high energies for which there is no clear supporting data at the available energies.
The relatively small uncertainty seen in the left panel of
Fig.~\ref{fig:omegapi_omegaK_opposite_signs} (compared to the uncertainties typically attached to the inelasticity, not employed in this work) reflects the uncertainty attached to the dispersive input $ \psi^0_0 (s) $.\footnote{As shown in Fig.~\ref{fig:phase_shift}, there are two compatible solutions for $ \psi^0_0 (s) $ (which on the other hand are not compatible for the function $|g^0_0 (s)|$ related to the inelasticity above the kaon pair threshold, which in any case does not appear in the analysis we now discuss) \cite{Pelaez:2020gnd}. In determining the error band seen in the left panel of Fig.~\ref{fig:omegapi_omegaK_opposite_signs}, we vary within $1\,\sigma$
the uncertainties attached to the parameters describing the $ \psi^0_0 (s) $ profile at high energies, which are relevant to determine the Omn\`{e}s solution at $M_D^2$, and also vary the speed with which the asymptotic behavior discussed in the text is reached, controlled by the parameter $m$ or $m^\ast$ introduced in Refs.~\cite{Moussallam:1999aq,Pich:2023kim}. By considering that the phase shift carries a small error,
one can show explicitly that the error band of $|\det \Omega^{(0)} (s)|$ tends to shrink
when $ |\det \Omega^{(0)} (s)| = 1 $ (i.e., when the integral in Eq.~\eqref{analyticalsol} vanishes), which happens at
$ s \simeq 2.1~\text{GeV}^2$.}

As previously mentioned, Eqs.~\eqref{eq:A0piSquared_constraint} and \eqref{eq:A0KSquared_constraint} are used in Ref.~\cite{Pich:2023kim} to select hadronic inputs, based on the $D \to \pi \pi$ and $D \to KK$ branching ratios, thus helping in the control of hadronic uncertainties; in particular, $ | T_{\pi\pi}^0 | \simeq |  T_{KK}^0| $ is then always obtained therein since any viable Muskhelishvili-Omn\`{e}s matrix solution must reproduce correctly the branching ratios.
Moreover,
Eqs.~\eqref{eq:sum_rule_rescattering} and \eqref{eq:DetOmega2_constraint} are automatically enforced in the method used in Ref.~\cite{Pich:2023kim}. Indeed, as also explained therein the explicit analytic expression of the determinant is used to specify the fundamental, or canonical, system of solutions to the dispersive integral equations.

The practical implication of Eqs.~\eqref{eq:A0piSquared_constraint}-\eqref{eq:DetOmega2_constraint}
can be more easily appreciated when $ | T_{\pi\pi}^0 | \simeq |  T_{KK}^0| $, observed to hold based on the isospin analysis of the experimental branching ratios, in which case
the expression for $ ( \omega^{({\rm Im})}_\pi - \omega^{({\rm Im})}_K )^2 $ satisfies:
\begin{equation}
    ( \omega^{({\rm Im})}_\pi - \omega^{({\rm Im})}_K )^2 \simeq | \det \Omega^{(0)} |^2 - \frac{1}{4} \left[ \frac{T_{\pi \pi}^{CC}}{T_{K K}^{CC}} \left( | \Omega^{(0)}_{11} |^2 - | \Omega^{(0)}_{21} |^2 \right) - \frac{T_{K K}^{CC}}{T_{\pi \pi}^{CC}} \left( | \Omega^{(0)}_{12} |^2 - | \Omega^{(0)}_{22} |^2 \right) \right]^2 \,,
\end{equation}
where $\lambda_s\approx -\lambda_d\approx |\lambda_d|$ has been used.
The later equation means that

\begin{equation}\label{eq:analytical_bound}
    ( \omega^{({\rm Im})}_\pi - \omega^{({\rm Im})}_K )^2 \lesssim | \det \Omega^{(0)} |^2 \,,
\end{equation}
which is notably simple, and in particular does not depend on the weak dynamics mediated by the quantities $T_{\pi \pi}^{CC}$ and $T_{K K}^{CC}$.
A way of achieving the maximum of $ ( \omega^{({\rm Im})}_\pi - \omega^{({\rm Im})}_K )^2 $ for fixed $| \det \Omega^{(0)} |$ (i.e., for fixed $ \psi^0_0 (s) $) is having $ | \Omega^{(0)}_{11} | \simeq | \Omega^{(0)}_{21} | $ and $ | \Omega^{(0)}_{12} | \simeq | \Omega^{(0)}_{22} | $ (which also depend on other independent, much more uncertain, dispersive inputs). In this case $ ( \omega^{({\rm Im})}_\pi - \omega^{({\rm Im})}_K )^2_{\rm max} \simeq | \det \Omega^{(0)} |^2 $, implying that the quantity on the left-hand side has a maximum defined uniquely by the profile of the experimental input $ \psi^0_0 (s) $. When $ \omega^{({\rm Im})}_\pi $ and $ \omega^{({\rm Im})}_K $ have opposite signs it is clear that their moduli are bounded; conversely, if they have the same sign, no bound can be set. As established by Eq.~\eqref{eq:sum_rule_rescattering}, the former case is realized in practice,
and after taking $ \omega^{({\rm Im})}_K $ negative we have that $ (\omega^{({\rm Im})}_\pi - \omega^{({\rm Im})}_K)_{\rm max} \simeq | \det \Omega^{(0)} | $. 
Considering negative values for $ \omega^{({\rm Im})}_K $, one has the plot shown in
the right panel of
Fig.~\ref{fig:omegapi_omegaK_opposite_signs}, where the inequality Eq.~\eqref{eq:analytical_bound} is shown as the blue region, and Eq.~\eqref{eq:sum_rule_rescattering} as a red dashed line chopped by the latter inequality. Namely, the acceptable values of the pair $ \{ \omega^{({\rm Im})}_\pi, \omega^{({\rm Im})}_K \} $ lay anywhere along the red dashed line.

We stress that the remarkable property of the approximate inequality Eq.~\eqref{eq:analytical_bound} is the dependence of its right-hand side on a reduced set of dispersive inputs:
$\det \Omega^{(0)}$ is thus a cleaner and more precise theoretical quantity than the individual elements of the Muskhelishvili-Omn\`{e}s matrix.
Indeed, in our previous work \cite{Pich:2023kim} we estimate the values of $ | \Omega^{(0)}_{ij} | $ and $ \phi_{ij} $
based on data from $ \pi \pi \to \pi \pi $, $ \pi K \to \pi K $, and $ \pi \pi \to K K $ rescattering.
In particular, the inelasticity function carries very large uncertainties for most energies above the kaon pair threshold, which thus reflect on the uncertainty in determining the individual elements of $\Omega^{(0)}$, as illustrated by their strong variability seen in Table~I of our previous work \cite{Pich:2023kim}.
Such inelasticity is irrelevant to the determination of $\det \Omega^{(0)}$.

Incidentally,
for the reference solution discussed in Ref.~\cite{Pich:2023kim}
the value for $ \omega^{({\rm Im})}_\pi - \omega^{({\rm Im})}_K $ saturates the bound of Eq.~\eqref{eq:analytical_bound} to a very good extent. It then follows that a combination of the pure isospin-$0$ contributions to the CP asymmetries of $ D^0 \to \pi \pi $ and $ D^0 \to K K $ is approximately maximized, in the sense that $ \omega^{({\rm Im})}_\pi - \omega^{({\rm Im})}_K $ is close to its maximum when varying dispersive inputs other than $ \psi^0_0 (s) $. We are then close to the black dot in the right panel of Fig.~\ref{fig:omegapi_omegaK_opposite_signs}.
Since they carry a relative minus sign as per Eq.~\eqref{eq:sum_rule_rescattering}, $ \omega^{({\rm Im})}_\pi $ and $ \omega^{({\rm Im})}_K $ are individually close to their maximum allowed values, and thus the same applies for the corresponding individual CP asymmetries in the isospin-$0$ case.

It is worth shifting the discussion toward more practical considerations.
What enters our current calculation is only the value of the determinant at the mass of the $D$ meson, $|\det \Omega^{(0)}(M_D^2)| \equiv |\det \Omega^{(0)}|$. We find that this presents small variations when the input phase $\psi^0_0(s)$ is varied within its uncertainties, namely $|\det \Omega^{(0)}|\in [0.28,0.34]$. Since we are here interested in setting bounds, we can conservatively take the highest value of $|\det \Omega^{(0)}|$ predicted, $|\det \Omega^{(0)}|=0.34$, which corresponds to an input $ \psi^0_0 (s) $ phase that goes faster to $360^\circ$
at high energies. It then follows that

\begin{eqnarray}
     |\omega_\pi^{\mathrm{(Im)}}| \lesssim  0.16 \,, \qquad |\omega_K^{\mathrm{(Im)}}| \lesssim  0.18 \,, \qquad \omega_\pi^{\mathrm{(Im)}}  \cdot  \omega_K^{\mathrm{(Im)}}<0 \,.
\end{eqnarray}

\noindent
Based on those values we get
\begin{equation}\label{eq:bounds_0_0}
 | a_{CP}(\pi\pi[0/0]) |\lesssim  2.7 \cdot 10^{-4} \,, \quad | a_{CP}(KK[0/0]) |\lesssim  1.4 \cdot 10^{-4} \,, \quad a_{CP}(\pi\pi[0/0]) \,\cdot\, a_{CP}(KK[0/0])<0 \,,
\end{equation}
and for the Belle~II observable \cite{Belle-II:2025rmf}:

\begin{equation}\label{eq:bound_BelleII}
   | R (0/0) | \lesssim 2.8 \cdot 10^{-4} \,,
\end{equation}

\noindent
where the large-$N_C$ counting is employed in deriving the numerical values in Eqs.~\eqref{eq:bounds_0_0} and \eqref{eq:bound_BelleII}.\footnote{Hereafter, the Wilson coefficients and quark masses are taken at 2~GeV.}
These are absolute upper bounds that we predict based on our method with the minimal use of scattering input.
The very stringent constraint on the CP violating observable $R (0/0)$, constructed out of the CP asymmetries in the $\pi^+\pi^-$ and $\pi^0\pi^0$ decays, is in sharp contrast with the currently measured large value of the $\pi^+\pi^-$ CP asymmetry.
Based only on our sum rule prediction in Eq.~\eqref{eq:bound_BelleII}, and taking the values of the measurements by LHCb \cite{LHCb:2019hro,LHCb:2022lry}, one expects an equally sizable as for $\pi^+\pi^-$ and opposite-sign CP asymmetry to appear on the $\pi^0\pi^0$ decay mode. However, as further discussed in the following, in Ref.~\cite{Pich:2023kim} we predicted levels of CP asymmetry much below the LHCb measurement.

\subsection{Unconstrained CP asymmetries}\label{sec:discussion_iso2_iso1}

We now discuss the remaining terms appearing in the decay CP asymmetries,
namely the asymmetries arising from the interference of $I=2$ ($1$) amplitudes with $I=0$ for the decays to a pion (respectively, kaon) pair, in view of Eqs.~\eqref{Tpipi2}-\eqref{twoamplitudesI0_kaons}.
Within our approach and keeping with the notation of our previous work \cite{Pich:2023kim}, we can express those asymmetries as 

\begin{eqnarray}\label{eq:aCP20_v1}
    a_{CP}(\pi\pi[2/0])&=&
 \frac{|T_{\pi\pi}^2|\mathcal{J}}{3\sqrt{2}\, |\lambda_d|\, |A(D^0\rightarrow \pi^+\pi^-)|^2} \nonumber \\
 &&\times \left[\bigl( T^{CC}_{K K}+  T^{P}_{K K} \bigr)|\Omega_{12}^{(0)}|\sin(\phi_2-\phi_{12}) 
 + T^{P}_{\pi\pi} |\Omega_{11}^{(0)}|\sin(\phi_2-\phi_{11}) \right]  ,
\end{eqnarray}
which for the values of the amplitudes extracted from the experimental branching ratios
is numerically 
\begin{equation}\label{eq:aCP20_v2}
     a_{CP}(\pi\pi[2/0])\,\simeq\, \left( 2.1 \,|\Omega_{11}^{(0)}|\,\sin(\phi_{11}-\phi_2)+3.2\, |\Omega_{12}^{(0)}|\,\sin(\phi_{12}-\phi_2)\right) \cdot 10^{-4} \, ,
\end{equation}
where $\phi_2$ is an effective $I=2$ strong phase.
Large-$N_C$ counting is used to derive Eq.~\eqref{eq:aCP20_v2} from Eq.~\eqref{eq:aCP20_v1}.

For the $I=1$ kaon pair information is scarce, in particular at the pertinent energies of the charm system; see, e.g., Ref.~\cite{Albaladejo:2015aca}.
Under the assumption of elasticity in $I=1$, we have
\begin{eqnarray}
\label{acpk1and0}
   A_{CP}(D^+\rightarrow K^+\overline{K}^0) &=& 0 \,, \\
   A_{CP}(D^0\rightarrow K^+K^-) &=& a_{CP}(KK[0/0])+ a_{CP}(KK[1/0]) \,, \\
   A_{CP}(D^0\rightarrow K^0\overline{K}^0) &=& \frac{|A[D^0\to K^+K^-]|^2}{|A[D^0\to K^0\overline{K}^0]|^2} \, \left\{ a_{CP}(KK[0/0])- a_{CP}(KK[1/0])\right\} \,. \label{eq:acpk0k0bar}
\end{eqnarray}
The decay asymmetry of $D^+$ to kaons vanishes up to kaon-mixing effects (for a recent discussion see, e.g., Ref.~\cite{Grossman:2026eit}).
The following sum rule can be written from Eq.~\eqref{eq:CPT_tilde}:

\begin{eqnarray}\label{4-channel-sumrule}
   && A_{CP}(K^+K^-) \,\mathcal{B}r (K^+K^-) + 2\, A_{CP}(K_S K_S)\,\mathcal{B}r (K_S K_S)\\
   &&  + \,
 2 \left( A_{CP}(\pi^+\pi^-) \,\mathcal{B}r (\pi^+\pi^-) + A_{CP}(\pi^0\pi^0) \,\mathcal{B}r (\pi^0\pi^0)\right)=0  \,, \nonumber
\end{eqnarray}
where the kinematical factors $\sigma_\pi$ and $\sigma_K$ have been absorbed into the definition of the branching ratios. In
each line the linear combination is such that only an overall 0/0 CP asymmetry remains.
Similar-form sum rules have been presented in the context of flavour $SU(3)$ \cite{Gronau:2005kz}.
A bound on the prediction for $a_{CP}(KK[0/0])$ is given in Eq.~\eqref{eq:bounds_0_0}. Also,
 
\begin{eqnarray}
     a_{CP}(KK[1/0])
     &= &- \frac{|T_{KK}^1|\, \mathcal{J}}{2|\lambda_s||A(D^0\rightarrow K^+K^-)|^2} \nonumber \\
     &&\times \left( T^{CC}_{\pi \pi} + T^{P}_{\pi \pi}+\frac{T^{CC}_{\pi\pi}}{T^{CC}_{KK}}T^{P}_{KK}\right)|\Omega_{21}|^{(0)}\sin(\phi_1-\phi_{21}) \,,
 \end{eqnarray}
which is numerically 
\begin{equation}\label{eq:numACP10-initial}
    a_{CP}(KK[1/0])\, \simeq \, 6\, |\Omega_{21}^{(0)}| \sin(\phi_1-\phi_{21}) \cdot 10^{-4} \, ,
\end{equation}
where large-$N_C$ counting has been used to obtain the numerical values. In particular, the kaon CP asymmetry can be simplified to the above expression because for the $I=1$ bare amplitudes, the relations $T^{CC}_{I=1,KK}=-T^{CC}_{KK}$ and $T^{P}_{I=1,KK}=-T^{P}_{KK}$ hold, where $T_{KK}^1 = \Omega^{(1)} \left( \lambda_s T^{CC}_{I=1,KK} - \lambda_b T^{P}_{I=1,KK} \right)$ is the full $I=1$ amplitude.
The phase $\phi_1$ is the strong phase induced by the final-state rescattering in the elastic $I=1$ $KK$ channel.
Small annihilation topology contributions have been neglected in the above expressions for kaons.

The rescattering parameters that those asymmetries depend on are all of the form $|\Omega_{ij}^{(0)}|\sin\delta'$, where $\delta'$ is a strong-phase difference. Within the currently discussed approach that only implements a minimal set of dispersive inputs, we cannot constrain those parameters. Let us however note the following: in order to amount to the level of $\pi^+\pi^-$ CP asymmetry experimentally observed close to $2\cdot 10^{-3}$ and taking into account the small allowed values for the other source of CP asymmetry $a_{CP}(\pi\pi[0/0])$, see Eq.~\eqref{eq:bounds_0_0}, one would need values of $|\Omega_{11}^{(0)}|$ and $|\Omega_{12}^{(0)}|$ around 4 at the most optimistic scenario of no destructive interferences. Comparing this value to the theoretically determined $|\det\Omega^{(0)}| \lesssim 0.4$ indicates that a high level of tuning would be needed. We also remind the reader that in our previous work \cite{Pich:2023kim}, where we have used the whole volume of strong-rescattering information, we do not find any apt Muskhelishvili-Omn\`{e}s matrix which would reproduce simultaneously the measured branching ratios and the CP asymmetries of LHCb, namely $|\Omega_{ij}^{(0)}|$ is found to be substantially below the unity.

\section{CP violating observables in the presence of NP}\label{section:NP}

In the following we examine how a potential heavy, weakly-coupled NP source would alter the CP-violating observables constructed in the previous sections. The parameterization in terms of isospin-invariant amplitudes $T_{\pi\pi}^{\{0,2\}}$ and $T_{KK}^{\{0,11,13\}}$ remains the same.\footnote{For the sake of simplicity we ignore isospin-invariant amplitudes from $\Delta I=5/2$ NP operators
in this discussion, which anyway do not appear in the dimension-6 weak effective Lagrangian in the presence of generic heavy NP.}
However, the presence of some NP dynamics can generate additional decay amplitudes with different weak phases, for instance through new contributions mediated by charged-current operators, or NP-mediated electroweak penguins that have been omitted in the basis of operators presented in Eq.~\eqref{eq:operator_list}. In the case of the $I=0$ channels, one can still without loss of generality split the isospin-invariant amplitudes into two amplitudes with different weak and potentially different strong phases, one of them CP-even and the second CP-odd, so expressions analogous to Eqs.~\eqref{twoamplitudesI0} and \eqref{twoamplitudesI0_kaons} would still hold true.
In the cases of $I=2$ and $1$ final states, the presence of NP can then generate additional amplitudes with a different weak phase $\text{Im} \{ \lambda_d \lambda_{NP}^\ast \} \neq 0$:
for the pion-pair state, Eq.~\eqref{Tpipi2} would then be modified to

\begin{equation}
    T_{\pi\pi}^2=\lambda_d \Tilde{A}_{d,\pi}^{(2)}+\lambda_{NP} \Tilde{A}_{NP,\pi}^{(2)} \,,
\end{equation}
with a similar equation holding for the $I=1$ kaon pair, namely

\begin{equation}
    T_{KK}^{13}=\lambda_d \Tilde{A}_{d,K}^{(13)}+\lambda_{NP} \Tilde{A}_{NP,K}^{(13)} \,.
\end{equation}
In principle then, a new CP asymmetry
could arise from the interference of these two amplitudes, namely a 2/2 (or 1/1) component, provided that they have different strong phases.

An important remark is in order for the cases of the 2/2 and 1/1 interferences and consequently for the charged-$D^\pm$ decay CP asymmetries. As known, a necessary ingredient for a non-vanishing CP asymmetry is the existence of two different strong phases. In the case of the charged-$D^\pm$ decay modes, there is only one possible isospin that the final states could be in. For the $I=2$ pion pair, based on the available data on strong scattering \cite{Durusoy:1973aj}, there is no clear indication of $I=2$ inelasticity
that could couple the two pions to other channels, at least up to the energies pertinent to charm systems.
The strong phases of the isospin-invariant $D$-decay amplitudes are then equal to the phase shift of the final-state scattering through Watson's theorem: 
\begin{equation}
    \arg \Tilde{A}^{(2)}_{d,\pi} = \arg \Tilde{A}^{(2)}_{NP,\pi} \mod 180^\circ = \arg(\pi\pi\to \pi\pi, \, I=2) \mod 180^\circ \,.
\end{equation}
Therefore, up to an overall sign, even a NP-induced amplitude
would carry the same strong phase as the SM amplitude, $\arg\Tilde{A}_{d,\pi}^{(2)}=\arg\Tilde{A}_{NP,\pi}^{(2)}$. As a result, the CP asymmetry of the $D^\pm$ decay modes is still expected to be very suppressed even if (heavy, weakly-coupled) NP is involved and independently of the rescattering dynamics in the $I=0$ block.
For the $I=1$ kaon pair
scattering data is scarce,
such that similar comments could hold for $I=1$.

\section{Conclusions}\label{sec:conclusions}

The main goal of this work is the discussion of direct CP violation in CP asymmetries of charm-meson decays to pion or kaon pairs.

We have reviewed the isospin decomposition of the decay amplitudes and indicated the different mechanisms contributing to the CP asymmetries in final states of definite electric charges. They stem from the interference of amplitudes of final states having total isospin-0 and -2 in the case of pion pairs, while in the SM the CP asymmetry from the interference of isospin-2 states is suppressed;
these different mechanisms of interference can be isolated with the $R (0/0)$ and $R (2/0)$ observables.

CP asymmetries in final states connected by rescattering are related due to unitarity and CPT.
Under the hypothesis of two-channel rescattering, we revisited a sum rule connecting the contributions to the CP asymmetries of pion and kaon pairs from the interference of isospin-0 final states. If further channels turn out to be relevant, such as a four-pion mode, then similar sizes of CP asymmetries would manifest therein.

Following our previous work \cite{Pich:2023kim},
we have factorized the flavour-changing electroweak transition from rescattering effects using large-$N_C$ counting.
We have then considered the analyticity of the rescattering part to write a dispersion relation, which requires knowledge of pion and kaon pair phase shifts and inelasticity functions produced by final-state interactions.
In Ref.~\cite{Pich:2023kim}, we considered the branching ratios to control the hadronic uncertainties stemming mainly from the inelasticity.
The main novelty of this work is the derivation of a relation among hadronic quantities that requires knowledge of a single dispersive input, namely the phase shift manifesting in $ \pi \pi \to K K $ scattering. Such a phase is precisely known and shown to lead to a precise bound on the difference of the hadronic quantities present in the CP asymmetries of isospin-0-with-0 interferences in pion and kaon pairs.
The CPT-unitarity sum rule was then used to further constrain the rescattering effect present in the CP asymmetries of isospin-0 final states.
Based on these considerations, it becomes more transparent that the level of CP asymmetry from the interference of isospin-0 amplitudes is much below the level of CP violation observed by LHCb, i.e., an order of magnitude smaller than the currently measured $\Delta A_{CP}^{\rm dir}$ and $D^0 \to \pi^+\pi^-$ CP asymmetry, and at the level of the $D^0 \to K^+K^-$ CP asymmetry.

On the other hand, the CP asymmetry from the interference of isospin-2 and isospin-0 states (or isospin-1 and isospin-0 states in the case of kaon pairs)
introduces new hadronic effects that can be determined from the available branching ratios.
Reproducing the level of CP violation found by LHCb, dominated by the 
$2/0$ CP asymmetry, requires a large amount of fine-tuning in the isospin-$0$ block in order to account simultaneously for the large CP asymmetry and the various branching ratios.
Based on the complete set of dispersive inputs studied in detail in Ref.~\cite{Pich:2023kim},
we found that
such a fine-tuned possibility is not favored,
with estimates for all CP asymmetries falling at the $10^{-4}$ order.

Finally, we argue that the decay mode $D^+\to\pi^+\pi^0$ is likely not an apt candidate for finding CP violation, even if some unknown beyond-the-SM dynamics is behind the current $\Delta A_{CP}^{\rm dir}$ puzzle.

\section*{Acknowledgements}

We thank I.~Danilkin, A.~Khodjamirian, B.~Kubis, B.~Moussallam, P.~Roig, and P.~Stoffer for useful discussions.
E.~S. would like to specially thank Tommaso Pajero for independently pointing out and discussing the role of experimentally used CP-asymmetry observables.
We acknowledge support by the Spanish Government (Agencia Estatal de Investigaci\'{o}n MCIN/AEI/10.13039/ 501100011033), Grants
No. PID2023-146220NB-I00 and CEX2023-001292-S,
and Swiss National Science Foundation (Project No. 10003620).

\bibliography{mybib}{}
\bibliographystyle{utphys}

\end{document}